# Denoising and Frequency Analysis of Noninvasive Magnetoencephalography Sensor Signals for Functional Brain Mapping

Abhisek Ukil, *Senior Member, IEEE*

*Abstract*—Magnetoencephalography (MEG) is an important noninvasive, nonhazardous technology for functional brain mapping, measuring the magnetic fields due to the intracellular neuronal current flow in the brain. However, most often, the inherent level of noise in the MEG sensor data collection process is large enough to obscure the signal(s) of interest. In this paper, a denoising technique based on the wavelet transform and the multiresolution signal decomposition technique along with thresholding is presented, substantiated by application results. Thereafter, different frequency analysis are performed on the denoised MEG signals to identify the major frequencies of the brain oscillations present in the denoised signals. Time-frequency plots (spectrograms) of the denoised signals are also provided.

*Index Terms*—Alpha wave, beta wave, brain signal processing, gamma wave, high beta wave, signal denoising, spectrogram, theta wave, threshold, time-frequency plot, wavelet transform.

## I. Introduction

MAGNETOENCEPHALOGRAPHY (MEG) is completely noninvasive, nonhazardous technology for functional brain mapping. Every current generates a magnetic field, and following the same principle in the nervous system, the longitudinal neuronal current flow generates an associated magnetic field. MEG detects weak extracranial magnetic fields in the brain, and allows determination of their intracranial sources [1], [2], giving a direct information on the brain activity, spontaneously or to a given stimulus. In comparison, computed tomography (CT) or magnetic resonance imaging (MRI) provides structural/anatomical information. Functional magnetic resonance imaging (fMRI), relying on blood flow, oxygenation changes, measures neuronal activity only indirectly [3]. By measuring these magnetic fields, scientists can accurately pinpoint the location of the cells/zones of the brain that produce each field. These spatio-temporal signals are used to study human cognition and, in clinical settings, for preoperative functional brain mapping, epilepsy diagnosis, etc. One common method of collecting functional data involves the presentation of a stimulus to a subject. However, most often, the inherent noise level in the data collection process is large enough to obscure the signal(s) of interest. In order to reduce the level of noise, the stimulus is repeated for as many as 100–500 trials. The trials are temporally aligned based on the timing of the stimulus presentation, and then an average is computed. This ubiquitously used approach works well, but it requires numerous trials. This, in turn, causes subject fatigue and, therefore, limits the number of conditions that can be tested for a given subject.

This paper presents a denoising algorithm using the wavelet transform along with thresholding, followed by different frequency analysis. The remainder of the paper is organized as follows. In Section II, information regarding the practical MEG technique, the associated noise problem, literature on denoising, and datasets used in this paper are provided. Section III provides a brief review of the wavelet transform. Section IV discusses in detail about the denoising technique and application results. Frequency analysis of the denoised signals in terms of sensor-specific brain oscillation frequency determination and time-frequency plots are described in Section V, followed by conclusions in Section VI.

## II. MEG, Noise, and Denoising

### A. MEG and Noise

The MEG technique measures the extremely weak magnetic field (of femto Tesla order, 1 fT = $10^{-15}$ T, compared to earth's magnetic field of 60 $\mu$T, about 9 orders of magnitude greater) generated by the intracellular neuronal current flow in the brain. This was initiated by the first recordings of the human magnetic alpha rhythm by Cohen in 1968 [2].

The spontaneous or evoked magnetic fields emanating from the brain induce a current in some induction coils, which produce a magnetic field in a superconducting quantum interference device (SQUID) [4]. Superconducting coils have essentially no electrical resistance; thus, the amount of current induced within the coil instantaneously tracks even very small changes in the magnitude of the impinging magnetic flux. The SQUID acts as a low-noise, high-gain, current-to-voltage converter that provides the system with sufficient sensitivity to detect neuromagnetic signals of fT order [5]. The MEG sensors consist of a flux transformer coupled to a SQUID, which amplifies the weak extracranial magnetic field and transforms it into a voltage. The sensors are immersed in liquid helium and attached on a concave bottom of a container, where they typically lie at a distance of 3–4 cm from the cortex [1]. In modern whole-head

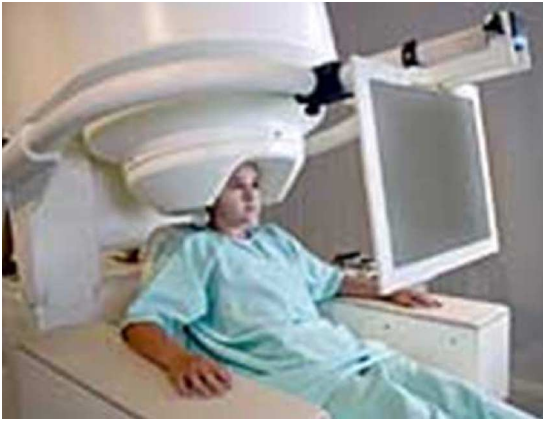

Fig. 1. Clinical use of MEG device with visual stimulus [6].

MEG devices [6], the patient wears a helmet containing 100+ sensors (see Fig. 1), in a costly magnetically shielded room. Visual (via screen) or audio stimulus is presented, followed by signal recording. MEG provides high-resolution measurement both in space (2–3 mm) and time (1 ms).

*B. Denoising Techniques*

Among different reported denoising techniques, independent component analysis (ICA)-based approaches are reported in [7] and [8], while time-shifted principal component analysis (PCA) is applied in [9]. Both ICA and PCA-based techniques depend on the time-domain data, not necessarily implying frequency-based thresholding like the wavelet analysis. The maximum-likelihood technique is used in [10], while blind source separation has been used in [11]. Both are promising, but the performance might vary for different datasets. Signal-space projection on source localization is used to reduce the artifact in [12], requiring higher-order computation. Adaptive filtering-based approaches like the adjusted least squares method [13], least mean square techniques [14] are also used. However, wavelet-based decomposition, especially the multiresolution analysis, would allow automatic adaptation. Time-delayed correlation-based techniques were used in [15] and [16], which are still heavily time-domain dependent. Probabilistic models have been used in [17] for noise suppression and source localization of MEG data. However, like [10] and [11], the performances based on probabilities might vary for different datasets.

*C. Experimental Setup and Dataset*

The dataset used in this paper was obtained from the competition on denoising of MEG data, organized by Hild in the 2006 IEEE Workshop on Machine Learning for Signal Processing (MLSP) [18].

The experimental setup used in this work consisted of 274 sensors detecting the magnetic field (fT) for the pre- and post-stimulus period, while the stimulus is presented to the subject at time $t = 0$ ms [18]. The subject was a 26–year-old right-handed female with no known mental deficits. The stimulus was provided using on/off pneumatic devices (having a compressed air-driven diaphragm). The total duration of the signal recording for each trial is for 361 ms, 120 ms for a pre- and 241 ms for a post-stimulus period [18]. The sampling frequency was 1.2 kHz [18]. We are interested in the denoising and analysis of the post-stimulus period. Ten trials of the MEG recorded signals using the aforementioned experimental setup have been used for the experimentation. Fig. 2 shows the average MEG signal for the 274 sensors over the ten trials.

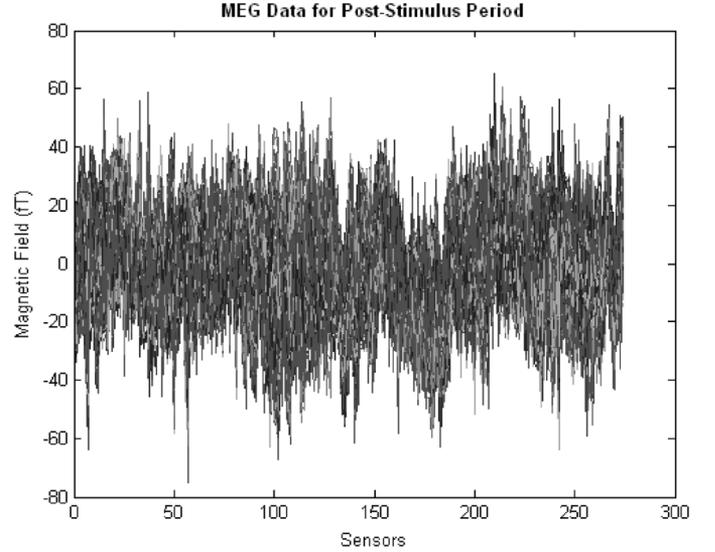

Fig. 2. Average MEG signal over ten trials for 274 sensors.

## III. WAVELET TRANSFORM

The wavelet transform (WT) is a mathematical tool, like Fourier transform for signal analysis. The WT breaks up a signal into shifted and scaled versions of the original (or mother) wavelet, allowing for simultaneous time and frequency analysis. In comparison, the Fourier transform uses sinusoids as the basis function, allowing only frequency analysis.

The continuous wavelet transform (CWT) [19] is defined as the sum over all time of the signal multiplied by the scaled and shifted versions of the wavelet function

$$\text{CWT}(a,b) = \int_{-\infty}^{\infty} x(t)\psi^*_{a,b}(t)\,dt \quad (1)$$

$$\psi_{a,b}(t) = |a|^{-1/2}\psi((t-b)/a). \quad (2)$$

$\psi(t)$ is the mother wavelet, the asterisk in (1) denotes a complex conjugate, and $a, b \in \Re, a \neq 0$ ($\Re$ is a real continuous number system) are the scaling and shifting parameters, respectively. $|a|^{-1/2}$ is the normalization value of $\psi_{a,b}(t)$ so that if $\psi(t)$ has a unit length, then its scaled version $\psi_{a,b}(t)$ also has a unit length.

The discrete wavelet transform (DWT) [19] is given by choosing $a = a_0^m, b = na_0^m b_0, t = kT$ in (1) and (2), where $T = 1.0$ and $k, m, n \in Z$ ($Z$ is the set of positive integers)

$$\text{DWT}(m,n) = a_0^{-m/2}\Big(\sum x[k]\psi^*[(k-na_0^m b_0)/a_0^m]\Big). \quad (3)$$

The multiresolution signal decomposition (MSD) [20], [21] technique decomposes a given signal into its detailed and smoothed versions. Let $x[n]$ be a discrete-time signal, then the MSD technique decomposes the signal in the form of WT coefficients at scale 1 into $c_1[n]$ and $d_1[n]$, the smoothed (time-domain view) and the detailed (frequency-domain view) coefficients, respectively

$$c_1[n] = \sum_k h[k - 2n]x[k] \qquad (4)$$

$$d_1[n] = \sum_k g[k - 2n]x[k] \qquad (5)$$

where $h[n]$ (lowpass) and $g[n]$ (highpass) are the associated filter coefficients that decompose $x[n]$ into $c_1[n]$ and $d_1[n]$, respectively. The decomposition process can be iterated, with successive approximations being decomposed in turn, so that the original signal is broken down into many lower resolution components. This is called the wavelet decomposition tree [20], [21], shown in Fig. 3. MSD technique can be realized with the cascaded quadrature mirror filter (QMF) [22] banks.

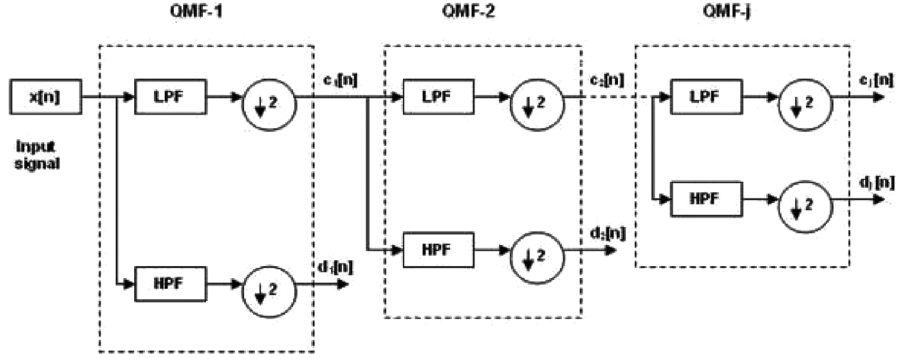

Fig. 3. Multiresolution signal decomposition.

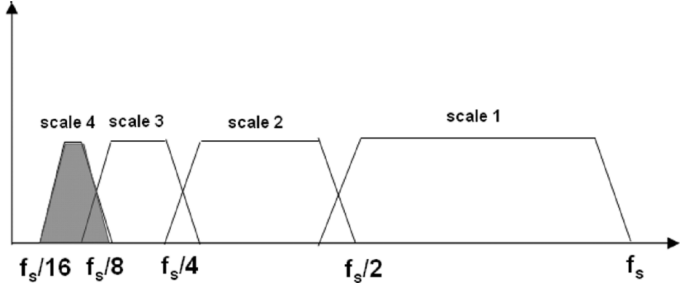

Fig. 4. Dyadic scales of sampling frequency associated with multiresolution decomposition.

## IV. Denoising

### A. Dyadic Decomposition

For the denoising purpose, the "db4" [19] mother wavelet is chosen. The sampling frequency of the recorded MEG signals was 1.2 kHz [18]. Utilization of the MSD technique implies decomposing the signal in dyadic scales of the sampling frequency, shown in Fig. 4. For this application, the MEG signals are decomposed up to four scales, which implies band-limitation of the signals up to $1200/2^4 = 75$ Hz.

The 274 sensor signals were decomposed individually up to four dyadic scales, each scale providing the smoothed ($c_1[n]$) and the detailed ($d_1[n]$) coefficients. So, the original signal $s$ can be represented as

$$s = c_4 + d_4 + d_3 + d_2 + d_1. \qquad (6)$$

With this scheme, if we are interested in reconstructing back the original signal, we combine $c_4 + d_4 = c_3$, and so on, to get back the original signal, i.e., traveling back in Fig. 3 of wavelet decomposition tree. An example decomposition using the post-stimulus signal from sensor 1 is shown in Fig. 5.

### B. Thresholding

Thresholding is the simplest method of signal separation or segmentation. Let $T$ denote the threshold. The hard thresholded signal $\hat{x}$ for any signal $x$ is

$$\begin{cases} \hat{x} = x, & \text{if } |x| > T \\ 0, & \text{otherwise}. \end{cases} \qquad (7)$$

Hard thresholding can be described as the usual process of setting to zero the elements whose absolute values are lower than the threshold. Soft thresholding [23] is an extension of hard thresholding, first setting to zero the elements whose absolute values are lower than the threshold, and then shrinking the nonzero coefficients towards zero.

After the dyadic decomposition, we apply the thresholding at each scale on the detailed coefficients, which represent the frequency profile of the signal in question. The key parameter in thresholding is obviously the choice of the threshold. For this application, "universal threshold" [24] is used.

The universal threshold $T$ is given by

$$T = \sigma\sqrt{2 \ln n} \qquad (8)$$

where $\sigma$ is the median absolute deviation of the wavelet coefficients, [24] and $n$ is the number of samples of the wavelet coefficients.

Fig. 6 shows the thresholding applied to the four-scaled decomposed signals (shown in Fig. 5) from sensor 1. The dotted horizontal lines indicate the thresholds at each scale.

In Fig. 6, adaptive thresholding technique is followed for the different scale. That is, for example, at scale 1, $d_1$ has frequency content from 300 to 600 Hz; therefore, a bigger threshold rejecting all these frequencies are selected compared to $d_3$, which has frequency content from 75 to 150 Hz. At each scale, part of the detailed coefficients that are less than the universal threshold is set to zero, and those that are greater than the threshold are shrinked towards zero. This soft-thresholding [24] is particularly effective in neuroinformatics application because the neuro signals are typically of lower frequency compared to high-frequency noise coming from the instrument, the environment, or other reasons.

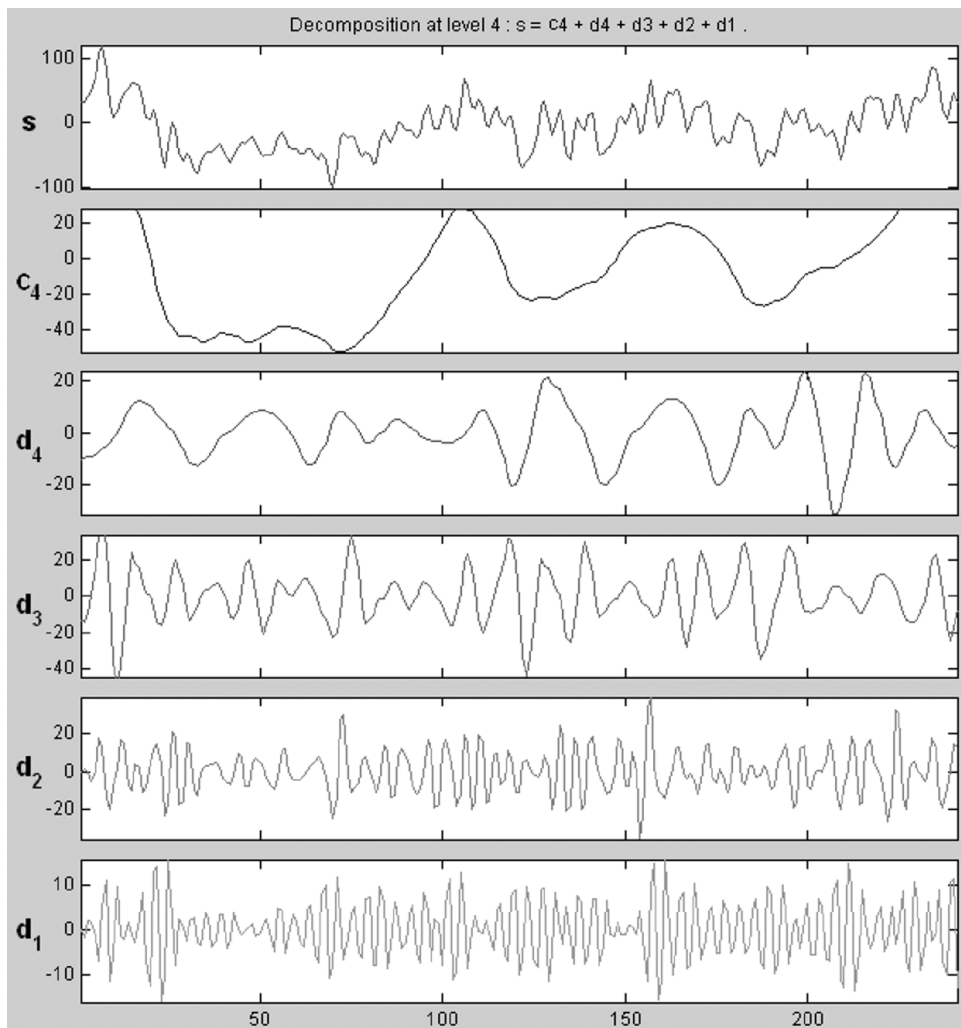

Fig. 5. Four-scale decomposition of sensor 1 signal.

### C. Results

After thresholding, the denoised detailed coefficients (scale 1 to 4) and the smoothed coefficients (scale 4) are combined as per (6) to get the denoised signal. The denoised version of the example signal (post-stimulus period) from sensor 1 is shown in Fig. 7. In Fig. 7, the blue (noisy) curve shows the original post-stimulus signal and the black (central) curve the denoised version. All the analysis was performed using the Matlab wavelet toolbox [25].

The denoising operation, i.e., wavelet decomposition followed by soft-thresholding using universal threshold, is performed for all the 274 sensors in this application. Fifteen examples of such denoising are shown in Fig. 9. These are chosen to representatively show the different brain oscillations (explained in Section V), in the original recordings vis-à-vis the denoised version. In all the plots of Fig. 9, the blue (noisy) curve shows the original post-stimulus signal, and the black (central) curve the denoised version.

### D. Performance

The performance metric used is the signal-to-noise/interference ratio (SNIR) [18]

$$\text{SNIR} \cong 10 \log \left[ \frac{1}{K} \sum_{i=1}^{K} \frac{\sum_{i=1}^{N} Y_{\text{mean}}^2}{\sum_{i=1}^{N} (Y_{\text{mean}} - Y_{\text{calc}})^2} \right] \text{ (dB)} \quad (9)$$

where $N = 241$ is the post-stimulus period in milliseconds, $K = 274$ is the number of sensors, $Y_{\text{mean}}$ is the average MEG signal computed over the ten trials (shown in Fig. 2), and $Y_{\text{calc}}$ is the denoised MEG signal.

For this application, the SNIRs achieved using different mother wavelets are summarized in Table I for the pre- and post-stimulus period.

### E. Comparative Analysis of Performance

From Table I, denoising using the db4 mother wavelet produces the best SNIR. It is to be noted that the dataset used here is the same used in the MLSP denoising competition [18].

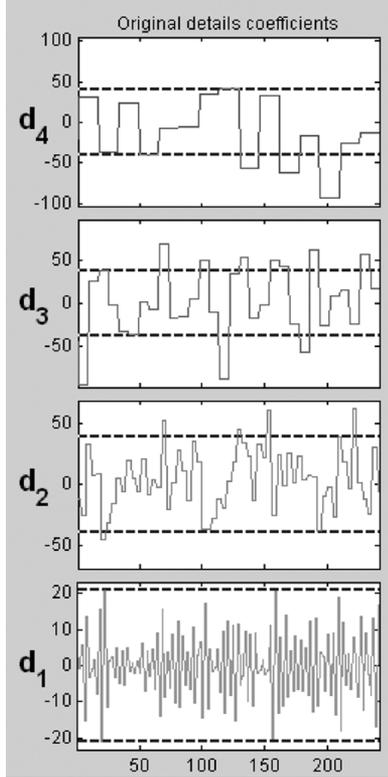

Fig. 6. Thresholding applied to the four-scale decomposition of sensor 1 signal.

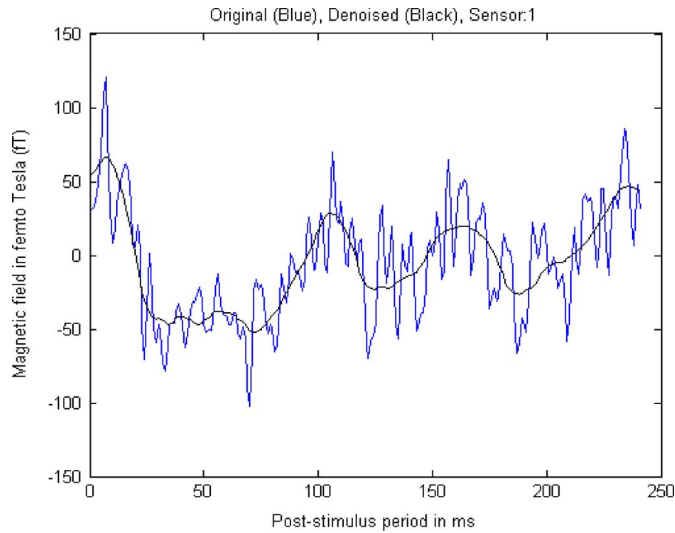

Fig. 7. Denoised version of the signal from sensor 1.

From the plots of the different wavelet functions in Fig. 8, the db10 wavelet function, with longer filter length, has a more oscillatory nature than the db4. For the low-frequency neuronal signals, therefore, db4 might yield a better cross-correlation than the db10, as demonstrated by the results. Nevertheless, both db4 and db10 being essentially unsymmetrical provides good performance. db10 has a sharper cutoff frequency. However, as we are not primarily interested in detecting discontinuities like changes in signal, this property is not of interest. coif1, with a more symmetrical wavelet function (see Fig. 8), performs a bit better than db10 for the pre-stimulus period. However, pre-stimulus period performance is of less interest than the post-stimulus period, the latter being more informative in terms of brain functionality.

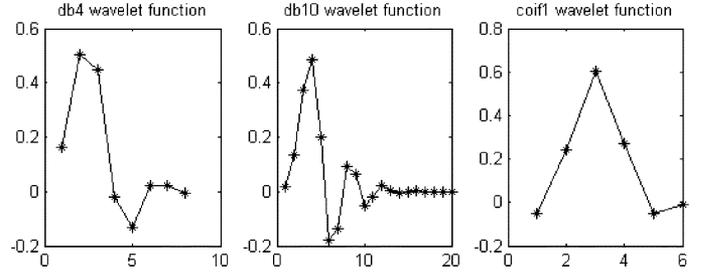

Fig. 8. Wavelet functions of the db4, db10, and coif1 mother wavelets.

Best reported SNIR using (9) for the denoised post-stimulus period in the MLSP 2006 competition was 0.6 dB [26], which used low-pass filtering with an averaging technique. The result obtained using the wavelet transform and thresholding in this paper surpasses that.

*F. Discussion of Results*

1) MEG signals are in general spatio-temporal low-frequency neuro signals which are strongly affected by instrumental and background noise, signals coming from the background brain activity, etc. Therefore, traditional denoising techniques (see Section II-B), especially filtering or time-shift-based approaches, fail to encompass the multitude of the potential sources of disturbances. On the other hand, the dyadic decomposition (see Fig. 4) capability of the DWT accounts for different sources of broadband noises, which makes it particularly suitable for denoising spatio-temporal MEG transient signals. This is also confirmed by the improved SNIR achieved by DWT-based approach in comparison with the low-pass filtering with averaging technique [26].

2) DWT decomposition combined with soft-thresholding is a useful approach as also demonstrated by Zhang and colleagues [27] for denoising Doppler ultrasound signals. That work considered db3, db4, and db6 mother wavelets. In the current work, other wavelets like Haar, coiflet, and symmlet are also compared with the Daubechies family like db4 and db10 [19] (Table I). Although the signals come from different sources in the current work and the work reported in [27], the common fact that db4 performs well to gain an SNIR of about 4 dB bolsters the potential of this type of mother wavelet and the overall approach in general. The work in [27], however, did not discuss about scales of decomposition, which is covered in details (see Section IV-A) in this paper. The choice of an appropriate scale of decomposition is important for optimally thresholding the noise, while preserving the signal of interest. This will be of particular help for subsequent processing as would be demonstrated for the MEG signals in the following sections.

3) The soft-, hard-, or combined thresholding is a useful approach because it is essentially an adaptive technique. This

TABLE I
PERFORMANCE (SNIR) OF DIFFERENT MOTHER WAVELETS FOR DENOISING PRE- AND POST-STIMULUS MEG SIGNALS

| Wavelet symbol | Wavelet name | Scale | SNIR (dB) for post-stimulus period | SNIR (dB) for pre-stimulus period |
|---|---|---|---|---|
| *Haar* | Haar | 4 | 3.8236 | 2.4253 |
| *db4* | Daubechies 4 | 4 | 4.3841 | 2.7804 |
| *db10* | Daubechies 10 | 4 | 4.3656 | 2.6592 |
| *coif1* | Coiflet 1 | 4 | 4.1845 | 2.6885 |
| *sym2* | Symmlet 2 | 4 | 4.2243 | 2.5523 |

means we do not necessarily need to have system knowledge *a priori*. This is a significant advantage over filtering and time-shifting-based approaches. Furthermore, this is not only useful for denoising MEG signals, but this approach is also used in other fields like power systems disturbance analysis [28].
4) The DWT and thresholding approach does not assume any Gaussian white noise type approximation for the affecting disturbances. Thereby, it is generally valid.
5) The applicability of the thresholding approach [24] for a 1-dimensional signal is exhaustively maintained by individually treating signals from each sensor. Significant denoising performance is achieved for the individual sensors (see Fig. 9). This also ensures avoidance of any ambiguity from the consideration that the signal is a linear combination of different sources or sensors [29]. Individual analysis of the sensors would also allow monitoring distribution of certain parameters of interest, e.g., frequency oscillations in the MEG signals, which is discussed in details in the following sections.

## V. FREQUENCY ANALYSIS

### A. Frequency-Domain Analysis

In this section, the basic frequency analysis of the denoised signals from the previous section alongside the original signals is presented. The original and the denoised signals were transformed into the frequency domain using the fast Fourier transform (FFT) [30], the sampling frequency being 1.2 kHz [18]. Then, for each of the 274 sensor signals, from the magnitude spectrum, the three maximum frequency contents are retrieved.

### B. Oscillations in Brain

Allocation of the sensor-wise maximum frequency by the aforementioned procedure leads us to the detection of the underlying brain waves. Distribution of the dominating frequencies with respect to how many sensors follow that is shown in Table II. From Table II, we can notice that primarily six types of waves are present, roughly of 5-, 10-, 15-, 20-, 25-, and 30-Hz frequencies. It is to be noted that the number of sensors having particular dominant frequency do not necessarily have to be physically adjacent. For example, from Fig. 9, spatially distributed sensor number 5, 44, 100, and 150 follow dominant 5-Hz oscillation; sensor number 2, 9, 232, and 264 follow dominant 10-Hz oscillation, etc.

TABLE II
DIFFERENT FREQUENCY WAVES

| Highest frequency (Hz) (denoised signal) | Number of sensors | Percentage |
|---|---|---|
| 4.98 | 149 | 54.38% |
| 9.96 | 49 | 17.88% |
| 14.94 | 47 | 17.15% |
| 19.92 | 20 | 7.30% |
| 24.90 | 4 | 1.46% |
| 29.88 | 5 | 1.82% |

Of these, the majority (54%) are in the range of 5 Hz. These are known as the "theta waves" [31], [32]. Theta waves involve many neurons firing synchronously, probably in the hippocampus and through the cortex [31]. Theta rhythms are normally absent in healthy awake adults but are physiological and natural in awake children under age 13. However, theta activity can be observed in adults during some sleep states, and in states of quiet focus, for example, meditation [33]. Besides, it has been widely acclaimed that theta waves are responsible for the "online" state of the hippocampus, one of readiness to process incoming signals [31]. Hasselmo showed that theta rhythms are very strong in rodent hippocampi and the entorhinal cortex during learning and memory retrieval [34], [35] and are believed to play a vital role in the cellular mechanism of learning and memory. This might also be the reason that the theta waves are dominantly present in this learning task-based MEG recording of adult subject (a 26-year-old right-handed female).

The next block is the 10-Hz frequency signals which are about 18%. These are known as the "alpha waves" [36]. Alpha waves are predominantly found to originate from the occipital lobe during periods of relaxation. These are electromagnetic oscillations in the frequency range of 8–12 Hz arising from synchronous and coherent electrical activity of thalamic pacemaker cells in the human brain.

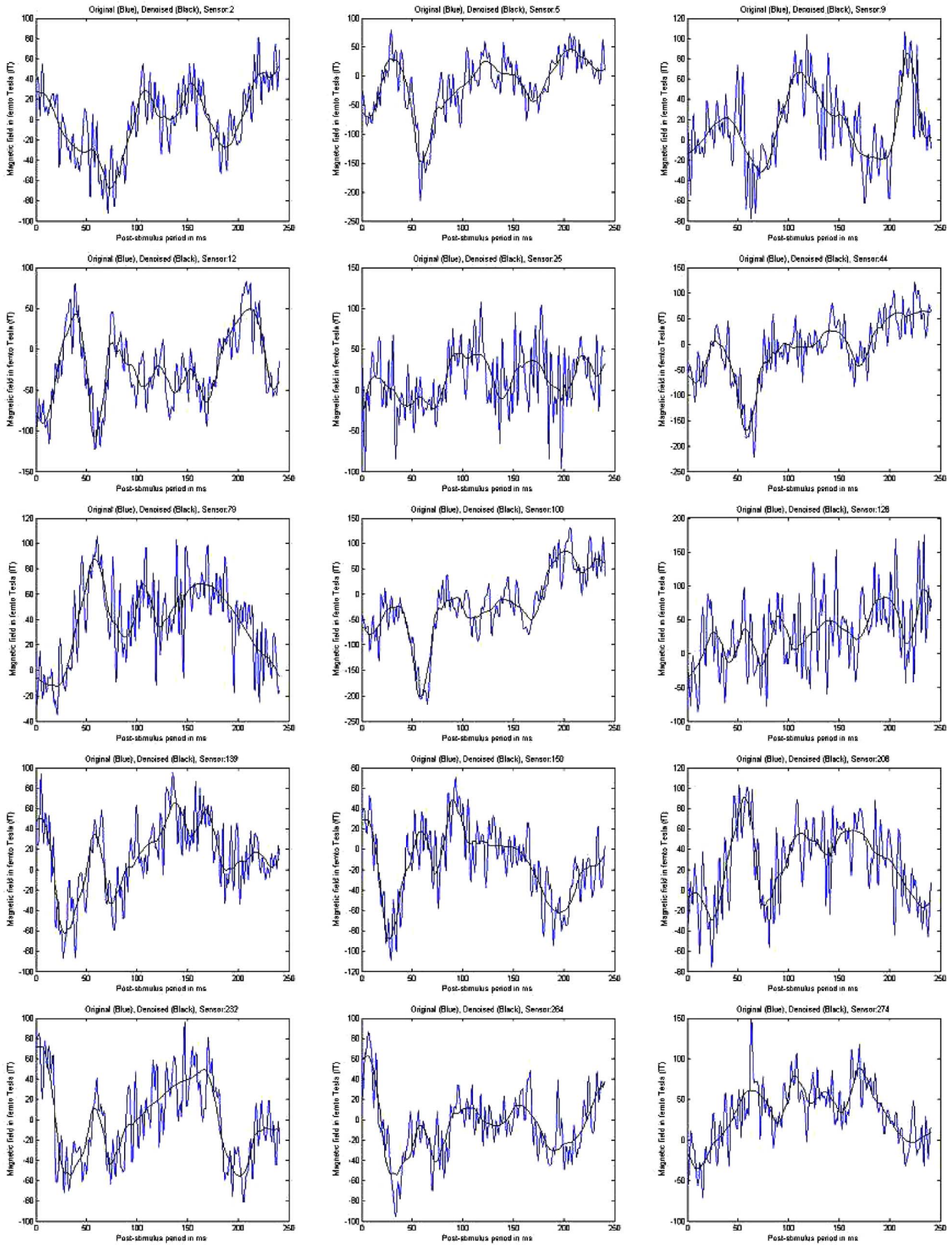

Fig. 9. Original post-stimulus (blue, noisy) and denoised (black, central) signals from different sensors.

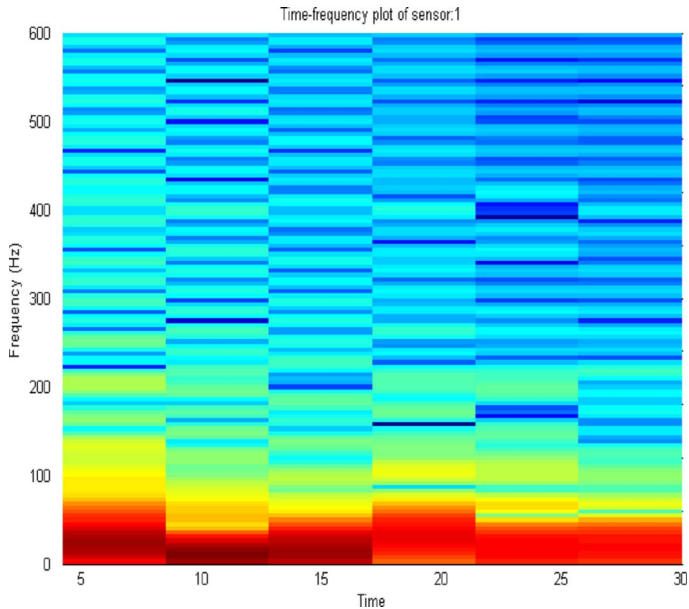

Fig. 10. Time-frequency plot (spectrogram) of the denoised sensor 1 signal.

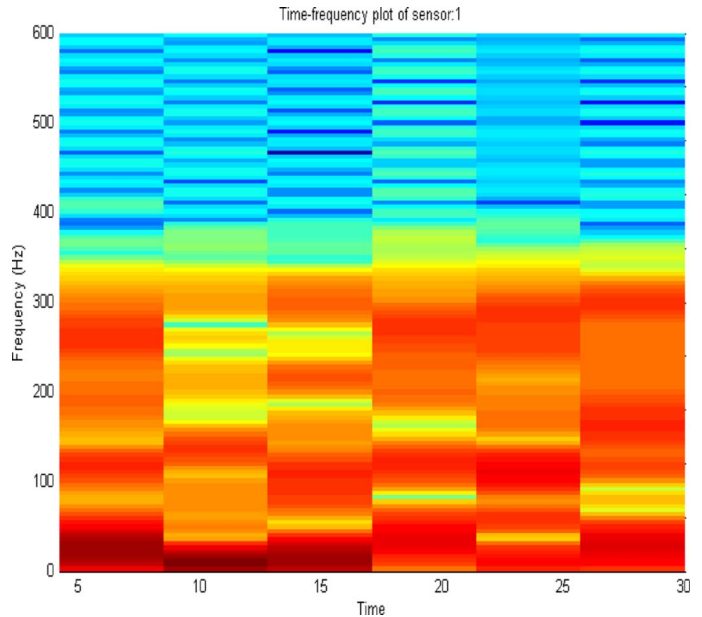

Fig. 11. Time-frequency plot (spectrogram) of the original noisy signal from sensor 1.

The next block 15 Hz (17%) is the "beta waves" [37]. Beta states are associated with normal consciousness in awake-state. Low amplitude beta with multiple and varying frequencies is often associated with active, busy, or anxious thinking and active concentration. The absence of beta waves indicates possible cortical damage of the subject. The 20-Hz frequency signals (7%) are also known as "high beta waves" [38].

The two other small blocks of 25 Hz (1.5%) and 30 Hz (2%) are towards the "gamma waves" [39]. Gamma oscillations, primarily generated from the dentate gyrus and CA3-CA1 regions, are believed to be responsible for various cognitive and motor functions [40].

For the purposes like functional analysis, source localization, etc., of the denoised MEG signals, distribution of the different frequency oscillations along with the known spatial positioning of the sensors could be of particular interest.

### C. Time-Frequency Plot: Spectrogram

Time-frequency plots are claimed to be very helpful by the neuroscientists in analyzing MEG signals. In time-frequency plot (also known as spectrogram), a sliding window of constant width is used over the whole length of the data, and the FFT is computed on the windowed data. This way, the short-time Fourier transform (STFT) [30] is calculated. The magnitude response of the STFT is plotted in the spectrogram in the way that the $X$ axis represents the sliding time, the $Y$ axis represents the frequency, and the color (intensity) represents the magnitude.

In Matlab implementation, by default the input signal is divided into eight segments [25]. If the signal cannot be divided exactly into eight segments, it is truncated [25]. The sampling frequency of 1.2 kHz [18] is used for estimating the frequency; otherwise, a normalized scale could be used [25]. Fig. 10 shows the time-frequency (spectrogram) plot for the denoised signal from sensor 1, and Fig. 11 shows the same for the original noisy signal from sensor 1. A comparison of Figs. 10 and 11 shows the effect of denoising, as in Fig. 11 we can notice the presence of lot of high-frequency noise compared to Fig. 10 containing mostly low-frequency signals of interest.

## VI. CONCLUSION

MEG, the noninvasive technique to measure the magnetic fields resulting from intracellular neuronal current flow, is quite important for functional brain imaging. However, the level of noise that is inherent in the data collection process is large enough that it oftentimes obscures the signal(s) of interest. Normal averaging over numerous trials of signal recording most often does not produce an optimum result and also causes subject fatigue. In this paper, the wavelet transform-based denoising technique of the MEG signal is presented. The MEG signals are first decomposed up to the fourth scale using multiresolution signal decomposition. This is followed by thresholding utilizing universal threshold. This method improves the SNIR up to about +4.38 dB compared to the reported +0.6 dB [26] on the same dataset.

After denoising, different frequency analyses are performed on the denoised signals. First, the sensor-specific frequency of the brain oscillations captured in the MEG signals are determined using the FFT. This helps in identifying the sensor-specific major brain oscillations, and localization of those with the availability of the spatial positioning information of the sensors. Furthermore, the STFT is computed by windowing the denoised MEG signals, and the magnitudes are plotted in conjunction with the frequency information and the sliding window time. These time-frequency plots (spectrograms) are often helpful in analyzing the MEG or similar signals by providing the frequency profile over the time-course.

### ACKNOWLEDGMENT

The author would like to thank K. E. Hild II for providing the recorded MEG signals and useful discussions.

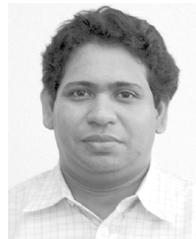

**Abhisek Ukil** (S'05–M'06–SM'10) received the B.E. degree in electrical engineering from the Jadavpur University, Calcutta, India, in 2000, the M.Sc. degree in electronic systems and engineering management from the University of Bolton, Bolton, U.K., in 2004, and the Ph.D. degree from the Tshwane (Pretoria) University of Technology, Pretoria, South Africa, in 2006, where he worked on signal processing and machine learning for power systems applications.

In 2006, he joined the ABB Corporate Research Center, Baden-Daettwil, Switzerland, where he is currently a Principal Scientist at the Integrated Sensor Systems Group. He is author/coauthor of more than 40 published scientific papers and the monograph *Intelligent Systems and Signal Processing in Power Engineering* (Springer, 2007) and is an inventor/co-inventor of six patents. His research interests include signal processing, machine learning, power systems, embedded systems, and computational neuro/bioscience.